# Construction of a Compact, Fully Automatic non-linear Absorption Spectrometer to Measure the Two-Photon Absorption Coefficient

H. Garcia,[1,a)] J. Serna,[2] and E. Rueda[3]

[1]*Physics department, Southern Illinois University, Edwardsville, Postal code, USA*

[2]*Grupo de Óptica y Espectroscopia, Centro de Ciencia Básica, Universidad Pontificia Bolivariana, Ca. 1 No. 70-01,*

*Campus Laureles, Medellín, Colombia*

[3]*Grupo de Óptica y Fotónica, Instituto de Física, U de A, Calle 70 No. 52-21, Medellín, Colombia*

Using an Electrically Focus Tunable Lens (EFTL), an integrating sphere and a tunable femtosecond-pulse laser (Mai Tai HP), we were able to measure the degenerate two-photon absorption coefficient (in transmission) of CdS and ZnSe in a long range of wavelengths (690-1040 nm), with a 5 nm resolution, in less than 30 minutes. We compared our results with theoretical approaches for the dispersion relations of the non-linear properties of semiconductors, and found excellent agreement with the experimental results. The system has no moving parts, is highly compact, and is fully automated.

Characterization of nonlinear optical properties of solids is based primarily on the use of experimental techniques that are based on single wavelength lasers, with only a few examples on the use of broadband lasers. Balu et al. [1–3] proposed two techniques to use the broadband spectrum from a femtosecond Ti:Saphire pulsed laser to measure non-linear absorption (NLA) and non-linear refraction (NLR) of solids: the first one uses a water cell to generate a broadband white-light continuum (WLC) ranging from 560 nm to 710 nm, and the second one uses high pressure Krypton gas. In both cases, to avoid the occurrence of nondegenerate nonlinearities, they used narrowband filters before the sample. Another approach to produce the WLC was proposed by Dey et al. [4] by using a photonic crystal fiber instead of the water cell. In all cases the techniques are based on the popular and straightforward Z-scan technique [5], where the NLA is obtained in transmission mode using an open aperture architecture and the mechanical motion of the sample through a high irradiance point. As pointed out recently by Steiger et al. [6], the problem with WLC systems is that they require complex optical paths. Instead, Steiger et al. used a broadband tunable femtosecond laser for the open aperture Z-scan setup, to measure the two-photon absorption (TPA) in photoinitiator materials. To achieve this, they emphasize the necessity of having, for each wavelength, all the experimental parameters (beam waist, laser power, pulse duration, etc.) fully characterized in order to avoid inaccuracies in the determination of the TPA values.

___________________________

a) Electronic mail: hgarcia@siue.edu.

In this work, we have accomplished this in an F-scan setup [7]. The broadband laser source (Spectra Physics Mai Tai HP, USA) is tunable from 690 nm – 1040 nm with a 1 nm resolution. Each of the wavelengths (690 nm- 1040 nm in steps of 5 nm) were fully characterized, the pulses' temporal widths were measured using an autocorrelation in a BBO crystal where angle phase matching produced a measurable signal or by looking at the pulses' spectral width and inferring from there the pulse duration assuming a transform limited pulse. Fig. 1 shows a typical autocorrelation trace and the corresponding spectral width at 788 nm. To measure the spot size at the focus for each of the wavelengths, the intensity of the beam was scanned with a 10 μm pinhole and modeled as a Gaussian beam profile. The Electrically Focus-Tunable lens (EFTL) (Optotune, EL10-30, Switzerland) generates the high and low irradiance points as described in [7]. In this setup, there is no need to characterize the EFTL because the measurement is based on high and low irradiance points that are used to extract the TPA coefficient. To eliminate the focusing and defocusing of the EFTL as the lens focal distance is changing, we used a set of photo-detectors (Thorlabs, In-FGAP71 150nm-550nm, Si-IDAS015 400nm-1100nm, and Ge-FD605 800nm-1800nm, USA) mounted in an integrating sphere (Newport, 819C-Sl-2, USA). In this way, the system can be adapted to work in a wider range of wavelengths.

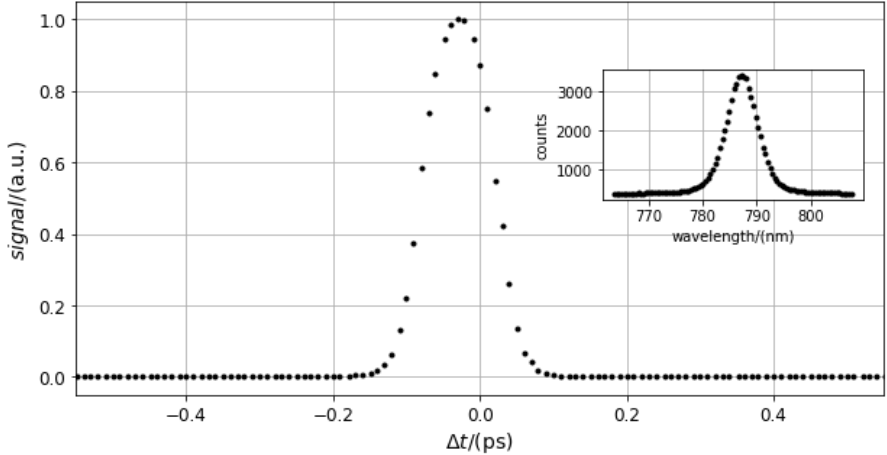

Fig. 1. Example autocorrelation of the Mai Tai tunable laser pulse; Δt is the time delay. (Inset) Spectral width.

Fig. 2 shows the experimental setup. The tunable laser generates on average 75 fs pulses at 80 MHz. The light passes through a continuous neutral density filter mounted on a computer controlled rotational stage, in a closed loop with a feedback mechanism, to maintain the average power of the laser at the sample almost constant (in our case at values ranging from 44 mW to 56 mW). After passing through the neutral density filter NF, the beam reaches the EFTL, focusing or defocusing it at the sample plane. The light transmitted through the sample is focused into the integrating sphere D1 by the lens L1, and corresponds to the NLA open aperture architecture. The signal is sent to the Lock-in amplifier that averages each of the data points 40 times to reduce the noise in the output. The wavelength of the laser is changed by the computer and the process is repeated until it reaches the maximum wavelength of the oscillator (1040 nm).

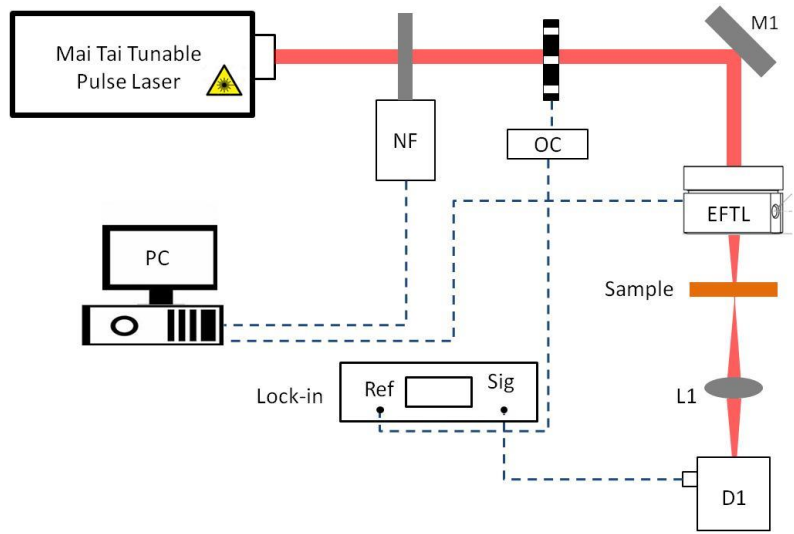

**FIG 2.** Experimental setup. The components are: Mai Tai tunable pulse laser, computer controlled motorized density filter (NF), Optical chopper(OC), Mirror (M1), Electrically focus-tunable lens (EFTL), sample, convergent lens (L1), integrating sphere (D1), Lock-in amplifier and personal computer (PC).

A typical trace for transmission is shown in Fig. 3(a). The position of the sample holder is such that minimum transmission takes place for a focal distance corresponding to 150 mA (programmed EFTL current for the case of 690 nm). The following paragraph describes the procedure of how the technique works and the experimental results for the case of two semiconductors, ZnSe and CdS.

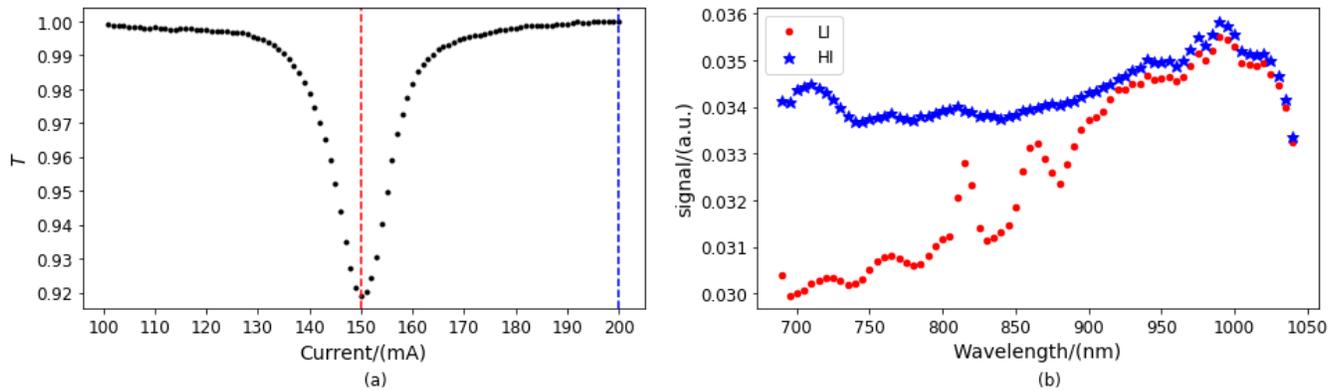

**Fig. 3.** (a) ZnSe F-scan signal at 690 nm. The red dashed line depicts the lowest transmittance (LI) at 150 mA, while the blue dashed line depicts the highest transmittance (HI) at 200 mA. Both values are used to determine the TPA coefficient at 690 nm. (b) HI and LI for every wavelength.

Because all the experimental parameters have to be well characterized in order to use the tunable laser, it is reasonable to think that for the case of Z-scan, in order to determine the TPA, for each wavelength only two values are needed, the highest transmittance $T_H$ at $z \to \infty$ and the lowest transmittance $T_L$ at $z = 0$. Thus, using the well-known expression for Z-scan open aperture architecture [8],

$$T(z) = 1 - \frac{\beta(1-R)I_0 L_{eff}}{2\sqrt{2}(1+x^2)} \qquad (1)$$

where $\beta$ is the TPA coefficient, $I_0$ is the maximum irradiance of the beam at $z = 0$, $R$ is the reflection coefficient at normal incidence, $L_{eff} = (1 - e^{-\alpha L})/\alpha$ is the effective length, α is the linear absorption, $L$ is the sample thickness, $x = z/z_0$, z is the position of the sample with respect to the beam waist, and $z_0$ is the Rayleigh range. By taking the highest and lowest transmittance, using Eq. (1), the relation $T_L/T_H$ will be equal to

$$\frac{T_L}{T_H} = 1 - \beta(1-R)I_0 L_{eff}/2\sqrt{2}, \qquad (2)$$

and the TPA coefficient for wavelength λ can be found from the relation

$$\beta(\lambda) = 2\sqrt{2}(1 - T_L/T_H)/(1-R)I_0 L_{eff} \qquad (3)$$

Although we use an F-scan setup instead of a Z-scan setup, because we only need the highest and lowest transmittance for each wavelength, it is true that $T_L/T_H = LI/HI$ (see Fig. 3(a)) and we can use Eq. (3) to determine the TPA. This is implemented experimentally by taking a trace where the current is change between 140 mA – 200 mA. After the trace is taken the system stores in memory the values for HI and LI. In Fig. 3(b) we have plotted the HI and LI for each wavelength.

From the above experimental analysis in Fig. 4 the value of $\beta$ as a function of energy is plotted for ZnSe and CdS, together with the theoretical prediction based on a two-band model [9] where the TPA coefficient is given by

$$\beta(\omega) = \frac{A}{E_g^3} \frac{\left(\frac{2\hbar\omega}{E_g} - 1\right)^{3/2}}{\left(\frac{2\hbar\omega}{E_g}\right)^5} H\left(\frac{2\hbar\omega}{E_g} - 1\right) \qquad (4)$$

A and $E_g$ have been used as fitting parameters and $H$ corresponds to the Heavyside step function. From the above model we have found an excellent agreement within the experimental error of the bandgap.

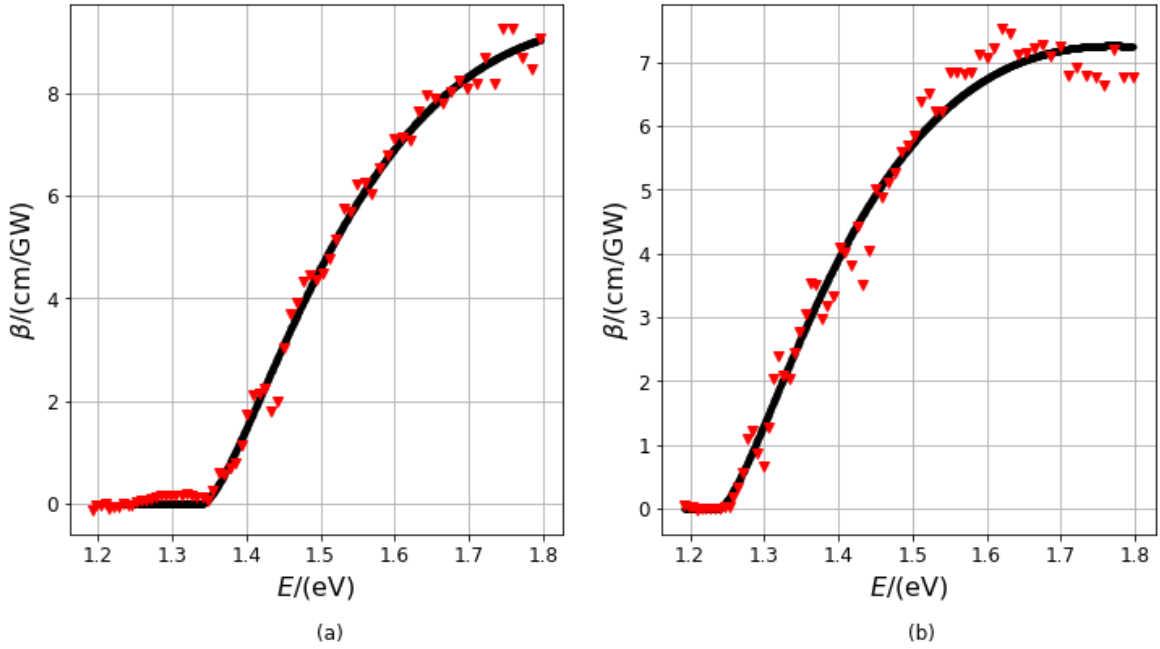

**Fig. 4.** (a) Two-photon absorption for ZnSe. The red triangles are the experimental data, and the black line is the fit obtained using the two-band model and parameters $E_g = 2.68$ eV and A = 3802. (b) Two-photon absorption for CdS. The red triangles are the experimental data, and the black line is the fit obtained using the two-band model and parameters $E_g = 2.47$ eV and A=2342. The statistical uncertainties of the measurements are under 0.05 cm/GW and are not plotted in the graphs.

The linear absorption for the two semiconductors was measured using an Ocean-Optics spectrometer (USB-400, USA) (see Fig. 5). The semiconductors bandgaps determined from Fig. 5 are 2.63 eV for ZnSe and 2.36 eV for CdS, which are in agreement with the values found with Eq. (4). It is worth mentioning that the TPA absorption can also be used to measure the bandgap of semiconductors, and this is of particular relevance to semiconductors where the direct gap transition may be forbidden by selection rules.

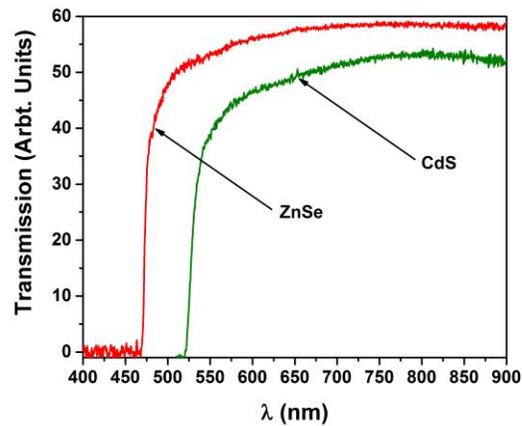

**Figure 5.** Transmittance for ZnSe ($E_g = 2.63$ eV), and CdS ($E_g = 2.36$ eV).

Because of the large range of the nonlinear TPA dispersion curve, in principle the nonlinear refraction $n_2$ can be obtained through a Kramers-Kronig relation [10] using the following equation

$$n_2(\omega) = \frac{c}{\pi} \int_{\omega_{min}}^{\omega_{max}} \frac{\beta(\omega')}{\omega'^2 - \omega^2} d\omega' + C \tag{5}$$

where $C$ is a constant that accounts for unknown contributions to the nonlinear refraction of the unknown TPA part of the spectrum, and that can be adjusted from already known values of $n_2$. This is an advantage as compared to the linear case where the absorption has to be known for a very large range of optical frequencies. In Fig. 6 we show the nonlinear refraction for ZnSe and CdS using Eq. (5) and the experimental data obtained for $\beta(\omega)$. Also, we show in solid black the theoretical data based on Eq. (5), using the values for $A$ and $E_g$ in Eq. (4). The agreement is remarkable when one takes into consideration that the contribution of the high frequency content is missing in the experimental data, which results in the two graphs deviating in this region.

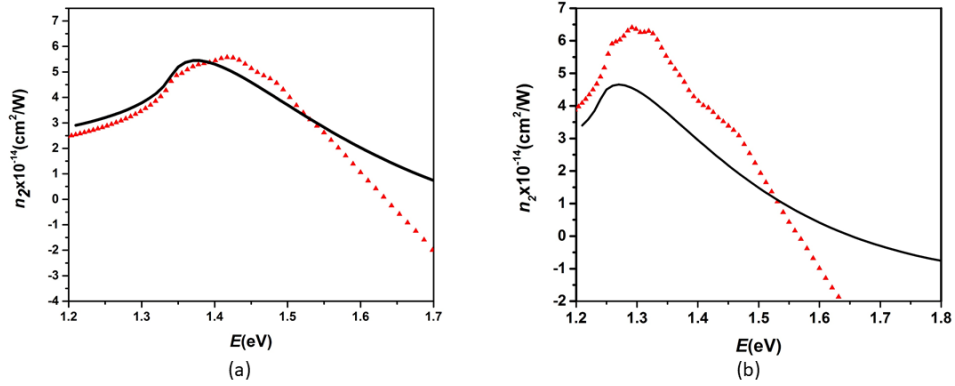

**Fig. 6.** Nonlinear refraction for (a) ZnSe and (b) CdS using the experimental data for $\beta(E)$ and Eq. (5) (red triangles), and using the $\beta(E)$ obtained from Eq. (4) and the values found for $A$ and $E_g$, and Eq. (5) (black line).

In conclusion, we have determined the two-photon absorption for CdS and ZnSe using a compact, fully automated system and with a broad spectral range (690 nm- 1040 nm) nonlinear absorption spectrometer. The system has no moving parts and each spectrum can be obtained in less than 30 minutes. The resolution of the system can be reduced down to 1 nm. One critical element is the use of an Electrically Focus Tunable Lens (EFTL) that eliminated the need for a translation stage reducing dramatically the vibrational noise typical in Z-scan. The results are in excellent agreement with the predictions of a two-band model and with the experimental value reported in the literature. We have also extended the analysis for the case of nonlinear refraction using Kramers-Kronig and found good agreement with the theoretical results taking into consideration that the theoretical model and the experimental data deviated at high frequencies. We are in the process of using the same system to measure the nonlinear refraction in an open aperture reflection mode for thin films where bulk TPA contributions can be neglected. The goal is to implement this system to measure the nonlinear absorption in transmission mode, and the nonlinear refraction in reflection mode for thin films, in this way the full nonlinear spectrum of the film can be obtained. This technique will open the doors for the development of a *nonlinear spectrometer*.

## ACKNOWLEDGMENTS

We acknowleged Dr. Jack Glassman for critcal reading of the manuscript. Dr. E. Rueda thanks Universidad de Antioquia (UdeA) for financial support. Dr. J. Serna acknowledges the support from Universidad Pontificia Bolivariana (UPB)., and Dr. H. Garcia thanks Southern Illinois University, Edwardsville (SIUE), for financial support.